\documentclass[conference]{IEEEtran}
\IEEEoverridecommandlockouts
\usepackage{cite}
\usepackage{array,amsmath,amssymb,amsfonts}
\usepackage{algorithmic}
\usepackage{graphicx}
\usepackage{textcomp}
\usepackage{xcolor}

\def\BibTeX{{\rm B\kern-.05em{\sc i\kern-.025em b}\kern-.08em
    T\kern-.1667em\lower.7ex\hbox{E}\kern-.125emX}}

\usepackage{adjustbox} 
\usepackage{multirow}
\usepackage{hyperref} 

\title{Privacy Threats and Countermeasures in Federated Learning for Internet of Things: A Systematic Review}

\author{
\IEEEauthorblockN{Adel ElZemity and Budi Arief}
\IEEEauthorblockA{School of Computing, University of Kent, Canterbury, United Kingdom \\
Email: ae455@kent.ac.uk, b.arief@kent.ac.uk}
}

\begin{document}

\maketitle

\begin{abstract}
Federated Learning (FL) in the Internet of Things (IoT) environments can enhance machine learning by utilising decentralised data, but at the same time, it might introduce significant privacy and security concerns due to the constrained nature of IoT devices. This represents a research challenge that we aim to address in this paper. We systematically analysed recent literature to identify privacy threats in FL within IoT environments, and evaluate the defensive measures that can be employed to mitigate these threats. Using a Systematic Literature Review (SLR) approach, we searched five publication databases (Scopus, IEEE Xplore, Wiley, ACM, and Science Direct), collating relevant papers published between 2017 and April 2024, a period which spans from the introduction of FL until now. Guided by the PRISMA protocol, we selected 49 papers to focus our systematic review on. We analysed these papers, paying special attention to the privacy threats and defensive measures -- specifically within the context of IoT -- using inclusion and exclusion criteria tailored to highlight recent advances and critical insights. We identified various privacy threats, including inference attacks, poisoning attacks, and eavesdropping, along with defensive measures such as Differential Privacy and Secure Multi-Party Computation. These defences were evaluated for their effectiveness in protecting privacy without compromising the functional integrity of FL in IoT settings. Our review underscores the necessity for robust and efficient privacy-preserving strategies tailored for IoT environments. Notably, there is a need for strategies against replay, evasion, and model stealing attacks. Exploring lightweight defensive measures and emerging technologies such as blockchain may help improve the privacy of FL in IoT, leading to the creation of FL models that can operate under variable network conditions. 
\end{abstract}


\section{Introduction}
The Internet of Things (IoT) consists of interconnected devices that communicate and exchange data, enhancing real-time data collection and analysis across sectors \cite{Madakam2015}. This connectivity introduces privacy and security challenges, necessitating solutions such as Federated Learning (FL) that train models on decentralised data. FL improves traditional machine learning by addressing issues of accuracy, efficiency, and privacy \cite{mcmahan2017communication}. However, FL in IoT faces challenges such as resource limitations, data heterogeneity, communication overheads, and privacy issues \cite{Aledhari_2020, 9475501}. These challenges are amplified by the limited computational power and energy resources of IoT devices, increasing potential risks to privacy~\cite{9767250}. Protecting FL data privacy on IoT devices is critical, and it requires robust defensive measures against threats such as inference attacks and data leakage. This review addresses the research gap by systematically analysing privacy threats and evaluating defensive measures within the IoT domain.

In order to address the identified research gap, this review systematically examines recent and pertinent literature. This review advances the knowledge regarding FL's applicability and privacy implications in IoT contexts by developing research questions centred on identifying privacy risks and defensive measures in such settings. 

Enhancing the taxonomy for FL privacy in IoT, the systematic classification of existing publications offers insights into privacy properties, potential threats, and defence mechanisms. The importance of this review stems from its comprehensive evaluation of FL's privacy implications in the IoT domain. It is a valuable resource for practitioners and researchers who aim to understand and manage the complex interactions between privacy and AI technologies in the IoT environments, which typically have very limited resources. Other published literature reviews tend to focus only on specific facets of FL or IoT privacy. In comparison, this review is notable for its thorough analysis of FL privacy in the IoT environments.

\vspace{0.1cm}
\noindent
\textbf{Contributions.} The key contributions of our paper are:
\begin{itemize}
    \item Comprehensive systematic review and analysis of privacy threats in Federated Learning (FL) within the Internet of Things (IoT) environments, including inference attacks, poisoning attacks, and eavesdropping.
    \item Evaluation of various defensive measures such as Differential Privacy and Secure Multi-Party Computation, assessing their effectiveness in safeguarding privacy without undermining the operational integrity of FL in IoT.
    \item Identification of critical research gaps, particularly highlighting the need for robust strategies against replay, evasion, and model stealing attacks, to enhance the privacy posture of FL in IoT.
\end{itemize}

The rest of the paper is organised as follows.  Section~\ref{background} introduces the important background of FL, especially in relation to privacy. Section~\ref{Scope of review} outlines our methodology, including a detailed explanation of the review's scope. Section~\ref{Results} presents our findings, while Section~\ref{Discussion} discusses the implications of these findings. Finally, Section~\ref{Conclusion} concludes our systematic review and suggests several areas for future research.

\section{Background}\label{background}
Privacy in the IoT domain faces complex challenges due to the ubiquitous nature of the devices involved, and the vast amount of data they collect. Privacy threats are exacerbated by the diversity and scale of IoT environments, making effective privacy protections crucial, yet difficult to achieve. Various studies highlight the need for robust privacy-preserving measures tailored to IoT’s unique constraints, such as device heterogeneity and extensive data generation~\cite{Tawalbeh2020IoT}. Moreover, emerging solutions need to focus on enhancing privacy without compromising the functionality and scalability of IoT systems~\cite{Qu2018Privacy}. 

Users of wearable and smart IoT devices are more worried than ever about how the personal data they collect is used and shared across services. Because of its volume and diversity, pervasive user data are beneficial for state-of-the-art machine learning and deep learning algorithms, which are being used in these applications more and more. To facilitate learning over a distributed network without transferring the data from each device, Federated Averaging (FedAvg)~\cite{Sun2021Decentralized} was presented as a foundational schema. FedAvg literature -- and the broader FL literature -- examine communication constraints and suggest enhanced learning frameworks, but do not investigate FL in severely constrained IoT environments with limited computing and storage capacity on the device~\cite{lo2021systematic}.

FL is a distributed machine learning technique where clients train locally without sharing personal data with the server~\cite{mcmahan2017communication}. Devices iteratively update a shared global model by aggregating information from each client model. Figure~\ref{fig:FL-Process} depicts the high-level architecture of the FL process, which usually consists of three phases~\cite{9475501}: 

\begin{figure}[!t]
    \centering
    \includegraphics[trim={0 1.5cm 0 1cm},clip,width=0.7\linewidth]{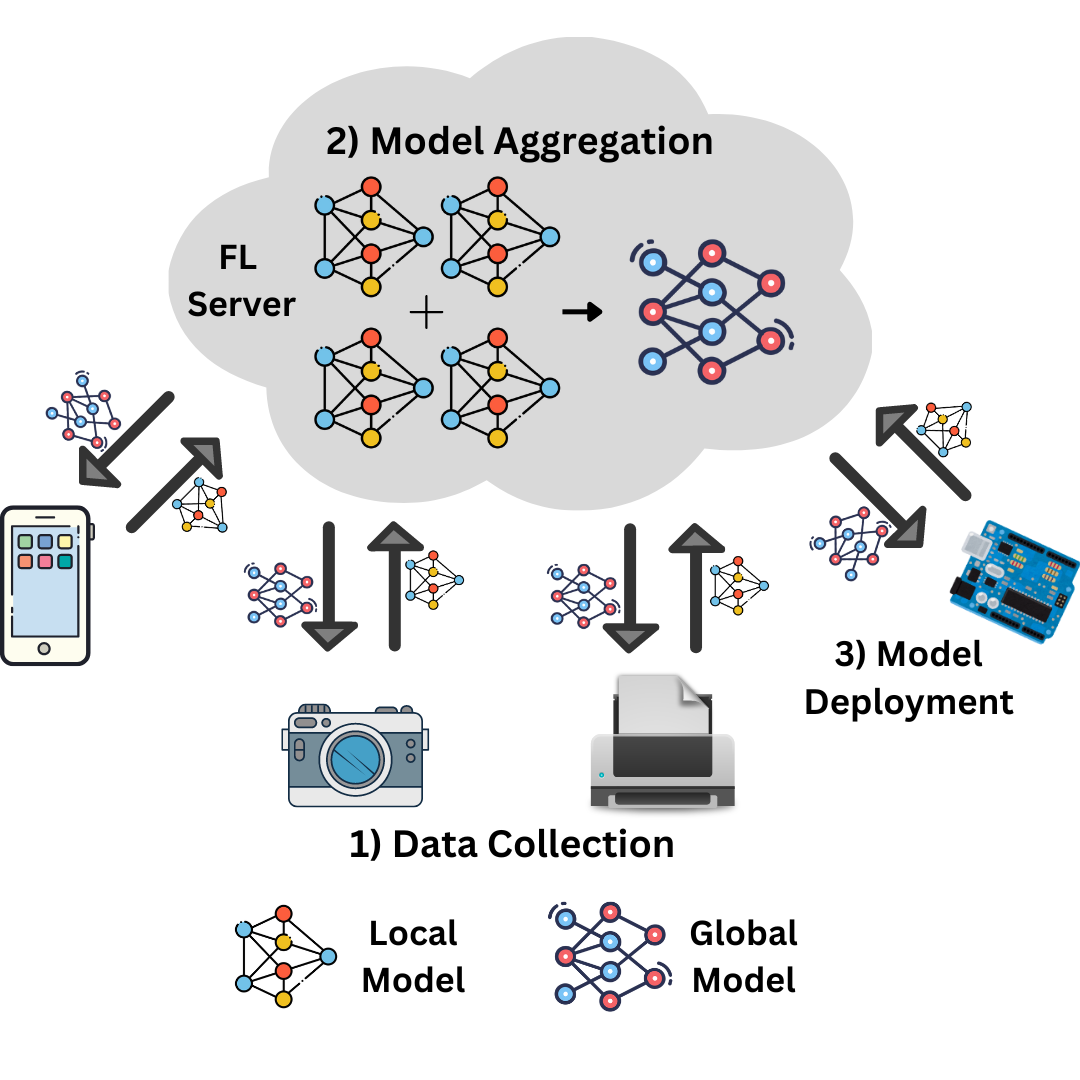}
    \caption{A High-level Architecture of FL Process}
    \label{fig:FL-Process}
\end{figure}

\begin{enumerate}
    \item \textbf{Data Collection and Local Model Update:} The target application and task requirements are determined by the central server during the first phase. The server initialises a global model (\(W_G^0\)) and transmits it to the chosen local clients, called participants. Every participant uses their local data to create a model. Each client $k$ updates its model parameters (\(W_t^i\)) to find the optimal parameters that minimise the local loss function (\(F_k(W_t^k)\)) after receiving the global model (\(W_G^t\)) (where $t$ denotes the $t$\textsuperscript{th} iteration). The local optimal models are then shared with the FL server.
    \item \textbf{Global Aggregation:} The FL server aggregates the local models provided by the participants to create an updated global model ($ W_G^{t+1} $). 
    \item \textbf{Model Deployment:} All of the new participants are given access to the most recent global model. Phases 2 and 3 are repeated until the central server reaches a convergence by minimising the global loss function (\(F(W_t^G)\)), which can be expressed as follows~\cite{li2020federated}: ($ \min _{w} f(w)=\sum _{k=1}^{N} P_{k} F_{k}(w) $) where $N$ is the total number of devices available, $F_k(w)$ is the expected prediction loss on a sample input of the $k$\textsuperscript{th} device on parameter $w$, $P_k(\ge 0)$ indicates the relative impact of each device $k$ while satisfying $\sum _{k} P_{k} = 1 $, and each device $k$ has $n_k$ samples (where $n= \sum _{k} n_{k} $). $P_k = (n_k/n)$ is the expression that can be used to show the relative impact of each local device.
\end{enumerate}

As this section has shown, current implementations of FL still face significant challenges, even though they offer promising paths for privacy-preserving ML, particularly within the IoT. These include protecting against sophisticated cyber threats that take advantage of the particular weaknesses of distributed architectures, managing resource constraints on IoT devices, and guaranteeing data privacy during model training. Significant gaps in privacy have come up from the inadequacies of existing strategies in effectively addressing these concerns, which our research attempts to address. The sections that follow will go into more detail about these issues and provide a new angle on privacy risks and the efficiency of modern defences in IoT environments.

\section{Methodology and Scope of Review}\label{Scope of review}
Our systematic literature review follows the PRISMA protocol \cite{moher2009preferred} for a rigorous and transparent approach. We defined research questions to identify privacy threats and defensive measures in FL within IoT contexts. Using a comprehensive search strategy and strict inclusion and exclusion criteria, we systematically analysed recent advances in the field.

To analyse the literature and compare the proposed techniques systematically, we established the following research questions to guide our assessment: 
        \begin{itemize}

        \item RQ1: What are the privacy threats present in federated learning within IoT environments?
        
        \item RQ2: What are the defensive measures to mitigate these risks without compromising data integrity, user privacy, and confidentiality?
        \end{itemize}

\subsection{Paper Selection and Data Collection}
We selected keywords aligned with our research questions, such as ``Federated Learning'', ``FL'', ``Decentralised Machine Learning'', ``Privacy-Preserving Machine Learning'', ``Resource'', ``Energy'', ``Power'', ``Limited'', ``Constrain'', ``Privacy'', and ``Threat''. The search query was: \textit{(``Federated Learning'' OR ``FL'' OR ``Decentralised Machine Learning'') AND (``IoT'' OR ``Internet of Things'') AND (``Resource'' OR ``Energy'' OR ``Power'') AND (``Limited'' OR ``Constrain'') AND (``Privacy'' OR ``Threat'')}. 

We used Scopus, IEEE Xplore, Wiley, ACM, and Science Direct to filter articles based on inclusion and exclusion criteria: articles from 2017 to April 2024, written in English, incorporating ``federated learning'' in the title, mentioning privacy aspects of FL or IoT, originating from reputable journals and conferences, and focusing on threats or defensive measures in IoT environments. Survey and review articles, and book chapters were excluded. Titles, abstracts, and full texts were evaluated against these criteria. Reference lists and citation tracking were used to ensure comprehensive coverage.

\subsection{Summary of Selected Papers}
Following the PRISMA protocol, \textbf{980} papers were identified through database searching. Additionally, we employed citation chaining, identifying \textbf{30} additional papers through backward and forward snowballing from our core articles to ensure thorough coverage of the literature. After removing duplicates, \textbf{970} papers remained, but \textbf{715} of which were then excluded after initial title and abstract filtering. Subsequently, the full-text of the remaining articles were subjected to further screening based on the inclusion and exclusion criteria by the researchers involved. In the event of disagreement between the researchers, a third researcher served as a mediator to resolve the selection conflict. Finally, \textbf{49} articles were selected for subsequent analysis in this systematic literature review.

\section{Results}\label{Results}
Existing reviews offer insights into the challenges and limitations of FL in IoT. Hosseinzadeh et al.~\cite{Hosseinzadeh2022Federated} discuss communication efficiency, resource allocation, and client selection in FL, focusing on its advantages without balancing potential drawbacks. Mothukuri et al.~\cite{Mothukuri2021A} highlight security threats such as communication bottlenecks and backdoor attacks in FL, but their review lacks comprehensive coverage of all privacy-related threats and limitations, particularly in resource-constrained IoT environments. Nguyen et al.~\cite{Nguyen2021Federated} emphasise security and privacy in FL for IoT networks but do not provide a comprehensive risk analysis. Similarly, Khan et al.~\cite{Khan2020Federated} discuss privacy challenges such as edge-cloud server inference and malicious user threats but lack an in-depth examination of IoT-specific vulnerabilities. Ferrag et al.~\cite{Ferrag2021Federated} focus on attack vectors such as model poisoning and inference attacks but do not extensively evaluate privacy challenges in the IoT context.

These reviews highlight the need for a comprehensive understanding of privacy concerns in FL within resource-constrained IoT environments. FL in IoT faces various threats across its phases. Table~\ref{threats_fed_learning} maps these threats to data collection, model aggregation, and model deployment phases, helping identify when specific threats are most likely to occur, which is crucial for developing targeted defences. For instance, inference attacks impact all phases, while model aggregation is most susceptible to various threats.

\begin{table}[!t]
\caption{Privacy Threats to Federated Learning Process}
\begin{center}
\begin{tabular}{|c|c|c|c|}
\hline
\multirow{2}{*}{\textbf{Threats}} & \textbf{Data } & \textbf{Model } & \textbf{Model } \\
& \textbf{Collection} & \textbf{Aggregation} & \textbf{Deployment} \\ 
\hline
Inference attacks & \checkmark & \checkmark & \checkmark \\
\hline
Poisoning attacks & \checkmark & \checkmark &  \\
\hline
Eavesdropping &  & \checkmark &  \\
\hline
Sybil attacks &  & \checkmark &  \\
\hline
Backdoor attacks &  & \checkmark & \checkmark \\
\hline
Gradient Leakage &  & \checkmark &  \\
\hline
Reconstruction &  &  & \checkmark \\
\hline
\end{tabular}
\label{threats_fed_learning}
\end{center}
\end{table}

\subsection{Threats}
A significant number of papers identify specific attacks that pose significant privacy risks and aim to validate their claims through proof of concept demonstrations. Subsequently, they propose various methods to defend against these identified threats. Table~\ref{attack_citations} groups the papers based on the seven privacy threats identified in the literature, and elaborated below.

\subsubsection{Inference Attacks}

\textbf{Membership inference attacks} pose a significant risk to users' privacy in resource-constrained IoT environments, determining if a specific data record was used in training a model. This can reveal sensitive information about the data subjects. Zhang et al.~\cite{Zhang2020GAN} discuss a membership inference attack using Generative Adversarial Networks (GANs) in FL, highlighting significant privacy leakages. This attack particularly affects FL models in IoT environments. Chen et al.~\cite{Chen2020Beyond} propose a novel user-level inference attack mechanism in FL, which is a critical concern for privacy in IoT implementations. Nguyen et al.~\cite{Nguyen2023Active} explore an active membership inference attack in FL under local differential privacy settings, demonstrating vulnerabilities in IoT data privacy. Zhao et al.~\cite{Zhao2021User-Level} analyse membership inference attacks at a user level within a FL framework deployed in a wireless IoT network. \textbf{Model inversion attacks} use model outputs to infer sensitive features of the input data. Salim et al.~\cite{Salim2022Perturbation-enabled} discuss FL's vulnerability to model inversion attacks in IoT-based social networks and propose a differential privacy-based framework to counter these threats. Xie et al.~\cite{Xie2021Privacy-Preserving} explore the challenges of resisting model inversion and extraction attacks in IoT using FL, proposing a lightweight privacy protection protocol for edge computing. Zhang et al.~\cite{Zhang2023Homomorphic} address privacy threats, including model inversion, using cryptographic methods within IoT-based healthcare systems employing FL. Zhou et al.~\cite{Zhou2020PrivacyPreserving} discuss protecting against model inversion attacks within a fog computing scenario using FL, focusing on the IoT context. \textbf{Property inference attacks} infer properties that hold over the entire training dataset or its subsets, which were not intended to be shared. A study by Shen et al.~\cite{Shen2021Exploiting} explores property inference attacks in blockchain-assisted FL within intelligent edge computing, specifically targeting unintended property leakages from model updates. Wang et al.~\cite{Wang2023Poisoning-Assisted} present novel methodologies for carrying out a poisoning-assisted property inference attack that specifically targets FL systems, aiming to infer properties of training data that are unrelated to the learning objective.

\begin{table}[t]
\caption{Reviewed Papers Grouped by Privacy Threats}
\vspace{-0.5cm}
\begin{center}
\begin{tabular}{|l|l|}
\hline
\textbf{Threat} & \textbf{Papers} \\
\hline
Inference Attacks & \cite{Zhang2020GAN}, \cite{Chen2020Beyond}, \cite{Nguyen2023Active}, \cite{Zhao2021User-Level}, \cite{Salim2022Perturbation-enabled}, \cite{Xie2021Privacy-Preserving}, \cite{Zhang2023Homomorphic}, \cite{Zhou2020PrivacyPreserving}, \cite{Shen2021Exploiting}, \cite{Wang2023Poisoning-Assisted} \\
\hline
Poisoning Attacks & \cite{Sun2020Data}, \cite{Li2023Multitentacle}, \cite{Zhang2021PoisonGAN}, \cite{Zhang2022RobustFL} \\
\hline
Eavesdropping & \cite{Zheng2022Exploring}, \cite{Ruzafa-Alcazar2023Intrusion}, \cite{Matheu2022Federated} \\
\hline
Sybil Attacks & \cite{Xiao2023SCA}, \cite{Jiang2021Sybil}, \cite{Fung2018Mitigating} \\
\hline
Backdoor Attacks & \cite{Hou2022Mitigating}, \cite{Ranjan2023Robust}, \cite{Yang2022Clean‐label}, \cite{Liu2023Facilitating} \\
\hline
Gradient Leakage & \cite{Zhu2023Defending} \\
\hline
Reconstruction & \cite{Li2021An}, \cite{Na2022Closing} \\
\hline
\end{tabular}
\vspace{-0.5cm}
\end{center}
\label{attack_citations}
\end{table}

\subsubsection{Poisoning Attacks} 
Adversaries intentionally manipulate the training data or the model updates to corrupt the learning process, leading to incorrect model outputs or leaking specific data characteristics. Sun et al.~\cite{Sun2020Data} discuss data poisoning attacks in FL within IoT systems, highlighting the vulnerability of federated models to such attacks and proposing a novel systems-aware optimisation method to derive optimal attack strategies. Li et al.~\cite{Li2023Multitentacle} explore adaptive poisoning attacks in the context of software-defined Industrial IoT (IIoT). They propose a framework that uses a tentacle distribution-based detection algorithm and a stochastic tentacle data exchanging protocol to minimise the impact of poisoned data. Zhang et al.~\cite{Zhang2021PoisonGAN} introduce PoisonGAN, a generative poisoning attack model for FL in edge computing. They demonstrate how this model can efficiently reduce attack assumptions and make attacks feasible in practice. Zhang et al.~\cite{Zhang2022RobustFL} propose RobustFL, a robust FL method for defending against poisoning attacks in IIoT, using an adversarial training framework. This method improves the resistance of the FL model to such attacks.

\subsubsection{Eavesdropping} Unauthorised interception of data during transmission between IoT devices and the central server or amongst the devices themselves, potentially exposing sensitive data. Zheng et al.~\cite{Zheng2022Exploring} explore FL as a method to preserve data training privacy from eavesdropping attacks in mobile-edge computing-based IoT. They propose a framework for optimising resource allocation to balance learning accuracy and energy consumption while protecting privacy. Ruzafa-Alcazar et al.~\cite{Ruzafa-Alcazar2023Intrusion} discuss the use of FL with differential privacy techniques to safeguard against intrusion and eavesdropping in IIoT environments. Matheu et al.~\cite{Matheu2022Federated} propose an FL approach to detect cyberattacks in IoT-enabled smart cities, integrating it with manufacturer usage descriptions to address eavesdropping and other attacks.

\subsubsection{Sybil Attacks} Attackers create multiple fake identities to influence the training process maliciously or to gain a disproportionate influence over the model. Xiao et al.~\cite{Xiao2023SCA} propose a novel approach for Sybil-based collusion attacks in IIoT FL systems, demonstrating how malicious participants can manipulate model aggregation through Sybil identities. Jiang et al.~\cite{Jiang2021Sybil} address Sybil attacks in the context of differential privacy-enhanced FL, proposing defence mechanisms that monitor training loss for anomalies to detect and mitigate such attacks. Fung et al.~\cite{Fung2018Mitigating} introduce ``FoolsGold'', a defence against Sybil-based poisoning attacks in FL, which identifies malicious Sybils by examining the diversity of client updates.  

\subsubsection{Backdoor Attacks}
Embedding hidden malicious functionality in the FL model, which can be activated to cause intended misbehaviour or to extract data. Hou et al.~\cite{Hou2022Mitigating} discuss a defence mechanism against backdoor attacks in IIoT applications using FL, incorporating federated backdoor filters with explainable AI models. Ranjan et al.~\cite{Ranjan2023Robust} propose graph-theoretic algorithms to identify and isolate backdoor attackers in FL systems, improving the robustness of the system. Yang et al.~\cite{Yang2022Clean‐label} explore clean-label poisoning attacks on FL in IoT environments, focusing on stealth and robustness of the attacks. Liu et al.~\cite{Liu2023Facilitating} enhance the effectiveness of early-stage backdoor attacks in FL by leveraging information leakage about the whole population's data distribution. 

\subsubsection{Gradient Leakage}
Even though raw data does not leave local devices, sharing model gradients can still leak information about the original data. Zhu et al.~\cite{Zhu2023Defending} focus on defending against inference attacks in FL within IoT, using parameter compression to mitigate the risk of gradient leakage.

\subsubsection{Reconstruction} Attackers use the gradients or model parameters shared during FL updates to reconstruct the inputs used in training. Techniques might involve solving optimisation problems that aim to find data points that would produce similar gradients. Li et al.~\cite{Li2021An} discuss the vulnerabilities of FL models to gradient-based reconstruction attacks, particularly in complex IoT environments. They propose a defence strategy suitable for resource-constrained IoT devices, emphasising adaptive communication to ensure model security and decrease communication overhead. Na et al.~\cite{Na2022Closing} reevaluate the effectiveness of current privacy-preserving techniques against reconstruction attacks in FL, proposing a new lightweight solution called Fragmented Federated Learning (FFL).

\subsection{Defensive Measures}
We also systematically evaluated the measures used within the literature to protect FL processes in IoT environments. 
Based on the specific privacy threats they address, we group seven defensive measures into three key categories: (i) Encryption and Obfuscation, (ii) Differential Privacy and Noise Injection, and (iii) Secure Multi-Party Computation and Anonymisation. These are detailed below.
Table~\ref{tab:defense_to_attack} provides a mapping of all defensive mechanisms against the types of threats they address. 
Related to this, Table~\ref{tab:quantitative_summary} provides a summary of quantitative metrics on various privacy threats and the effectiveness of corresponding defensive measures, showing the metrics before and after applying these defensive measures.

\begin{table*}[t]
\caption{Mapping of Privacy Defensive Measures Against Privacy Threats}
\begin{center}
\begin{tabular}{|l|c|c|c|c|c|c|c|}
\hline
& \rotatebox{90}{\textbf{Inference Attacks}} & \rotatebox{90}{\textbf{Poisoning Attacks}} & \rotatebox{90}{\textbf{Eavesdropping}} & \rotatebox{90}{\textbf{Sybil Attacks}} & \rotatebox{90}{\textbf{Backdoor Attacks}} & \rotatebox{90}{\textbf{Gradient Leakage}} & \rotatebox{90}{\textbf{Reconstruction}} \\
\hline \hline
\textbf{Gradient Obfuscation} & \cite{Wu2022Defense} & & & & & & \\
\hline
\textbf{Parameter Compression} & \cite{Zhu2023Defending}, \cite{Chen2023Efficient} & & & & & & \\
\hline
\textbf{Compressed Sensing} & \cite{Miao2023Efficient} & & & & \cite{Li2021Communication} & & \\
\hline \hline
\textbf{Differential Privacy} & \cite{Shen2022Performance}, \cite{Cui2021Security}, \cite{Ruzafa-Alcazar2023Intrusion}, \cite{Yin2021Privacy}, \cite{He2024Clustered} & \cite{Shen2022Performance}, \cite{Cui2021Security}, \cite{Ruzafa-Alcazar2023Intrusion}, \cite{Yin2021Privacy}, \cite{He2024Clustered} & & & & & \\
\hline
\textbf{Decentralised Perturbation} & \cite{Arachchige2020Privacy}, \cite{Mothukuri2021Federated}, \cite{Mantey2023Blockchain}, \cite{Alotaibi2021Biserial}, \cite{Alamleh2022Federated} & & & \cite{Mothukuri2021Federated} & & & \\
\hline \hline
\textbf{Secure Multi-Party Computation} & \cite{Li2020Privacy-Preserving}, \cite{Liu2023Comprehensive}, \cite{Lu2020Blockchain}, \cite{Zhou2020PrivacyPreserving} & \cite{Li2020Privacy-Preserving}, \cite{Liu2023Comprehensive}, \cite{Lu2020Blockchain}, \cite{Zhou2020PrivacyPreserving} & \cite{Gade2018} & & \cite{Liu2023Comprehensive} & \cite{Fu2020} & \\
\hline
\textbf{Anonymisation and Siamese Networks} & \cite{Song2020Analyzing} & & & \cite{Song2020Analyzing} & & & \cite{Yue2022} \\
\hline
\end{tabular}
\end{center}
\label{tab:defense_to_attack}
\end{table*}

\begin{table*}[t!]
\caption{Summary of Quantitative Metrics on Privacy Threats and Defensive Measures in Federated Learning}
\centering
\begin{tabular}{|m{1.1cm}|m{1.8cm}|m{2.2cm}|m{4.9cm}|m{5.9cm}|}
\hline
\textbf{Paper} & \textbf{Attack Type} & \textbf{Defence Measure} & \textbf{Metrics Before Defence} & \textbf{Metrics After Defence} \\
\hline
Zhu et al. (2023) \cite{Zhu2023Defending} & GAN-Based Privacy Inference & FLPC$^{\mathrm{a}}$ & 
\textbf{Accuracy:} 0.9591 $\rightarrow$ 0.9507 (Baseline) \newline Decrease: 0.84\% & 
\textbf{Accuracy:} 0.9565 $\rightarrow$ 0.9557 (FLPC, Comlevel = 0.001) \newline Decrease: 0.08\% \\
\hline
Song et al. (2020) \cite{Song2020Analyzing } & User-Level Privacy Attack & mGAN-AI$^{\mathrm{b}}$ & 
\textbf{Accuracy (MNIST Training):} 0.9438 \newline 
\textbf{Accuracy (MNIST Testing):} 0.9247 \newline 
\textbf{Accuracy (AT\&T Training):} 0.9435 \newline 
\textbf{Accuracy (AT\&T Testing):} 0.9267 & 
\textbf{Passive mGAN-AI Inception Score:} 1.42$\pm$0.02 \newline 
\textbf{Active mGAN-AI Inception Score:} 1.61$\pm$0.05 \newline 
\textbf{Passive mGAN-AI Accuracy:} Similar to baseline \newline 
\textbf{Active mGAN-AI Accuracy:} Slightly lower \\
\hline
Liu et al. (2021) \cite{liu2021privacy} & Label-Flipping & PEFL$^{\mathrm{c}}$ & 
\textbf{Attack Success Rate:} 0.001 $\rightarrow$ 1 \newline 
\textbf{True Positive Rate:} 0.98 $\rightarrow$ 0 \newline 
\textbf{Non-Source Class Accuracy:} 0.95 $\rightarrow$ 0.51 & 
\textbf{Attack Success Rate:} 0 $\rightarrow$ 0.03 \newline 
\textbf{True Positive Rate:} 0.95 $\rightarrow$ 0.88 \newline 
\textbf{Non-Source Class Accuracy:} 0.97 $\rightarrow$ 0.76 \\
\cline{2-5}
& Backdoor & PEFL$^{\mathrm{c}}$ & 
\textbf{Attack Success Rate:} 0.001 $\rightarrow$ 1 \newline 
\textbf{True Positive Rate:} 0.98 $\rightarrow$ 0 \newline 
\textbf{Non-Source Class Accuracy:} 0.95 $\rightarrow$ 0.64 & 
\textbf{Attack Success Rate:} 0 $\rightarrow$ 0.04 \newline 
\textbf{True Positive Rate:} 0.95 $\rightarrow$ 0.88 \newline 
\textbf{Non-Source Class Accuracy:} 0.97 $\rightarrow$ 0.76 \\
\hline
Liu et al. (2023) \cite{liu2023privacy} & Gradient-Based Data Reconstruction & Privacy-Encoded FL & 
\textbf{PSNR:} 28.99 $\rightarrow$ 0.6838 (Baseline) \newline 
\textbf{Test Accuracy:} 89.76\% & 
\textbf{PSNR:} 29.45 $\rightarrow$ 3.54 \newline 
\textbf{Test Accuracy:} 87.78\% $\rightarrow$ 89.86\% \\
\hline
Jiang et al. (2020) \cite{jiang2020mitigating} & Sybil Attacks & DP$^{\mathrm{d}}$ and Anomaly Detection & 
\textbf{CNN Error Rate:} 0.03 $\rightarrow$ 0.14 \newline 
\textbf{MLP Error Rate:} 0.59 $\rightarrow$ 0.63 & 
\textbf{CNN Error Rate:} 0.03 $\rightarrow$ 0.03 \newline 
\textbf{MLP Error Rate:} 0.59 $\rightarrow$ 0.59 \\
\hline
Miao et al. (2022) \cite{miao2022against} & Backdoor Attacks & CND$^{\mathrm{e}}$ with DP$^{\mathrm{d}}$ & 
\textbf{CIFAR-10 Accuracy:} 88\% $\rightarrow$ 84\% \newline 
\textbf{EMNIST Accuracy:} 99\% $\rightarrow$ 90\% \newline 
\textbf{CIFAR-10 Attack Success:} 0\% $\rightarrow$ 80\% \newline 
\textbf{EMNIST Attack Success:} 0\% $\rightarrow$ 100\% & 
\textbf{CIFAR-10 Accuracy:} 82\% $\rightarrow$ 81\% \newline 
\textbf{EMNIST Accuracy:} 95\% $\rightarrow$ 75\% \newline 
\textbf{CIFAR-10 Attack Success:} 0\% $\rightarrow$ 3\% \newline 
\textbf{EMNIST Attack Success:} 0\% $\rightarrow$ 5\% \\
\hline
Li et al. (2023) \cite{li2023privacy}  & Gradient-Based Inference, Byzantine & PBA$^{\mathrm{f}}$ & 
\textbf{Global Accuracy:} 88\% & 
\textbf{GA$^{\mathrm{g}}$ (f=2):} $\sim$87\%; 
\textbf{GA$^{\mathrm{g}}$ (f=6):} $\sim$83\% \newline 
\textbf{LFA$^{\mathrm{h}}$ (f=2):} $\sim$85\%; 
\textbf{LFA$^{\mathrm{h}}$ (f=6):} $\sim$60\% \newline 
\textbf{Running Time:} 9.391 s \\
\hline
Asad et al. (2020) \cite{asad2020critical} & DP$^{\mathrm{d}}$, Homomorphic Encryption, Backdoor & DP$^{\mathrm{d}}$, HE$^{\mathrm{i}}$, Secure Aggregation & 
\textbf{DP Accuracy (PB$^{\mathrm{j}}$=0.1):} 70\% \newline 
\textbf{DP Accuracy (PB$^{\mathrm{j}}$=0.5):} 55\% \newline 
\textbf{DP Accuracy (PB$^{\mathrm{j}}$=1.0):} 40\% \newline 
\textbf{DP Accuracy (PB$^{\mathrm{j}}$=2.0):} 30\% \newline 
\textbf{HE Accuracy (SP$^{\mathrm{k}}$=32):} 85\% \newline 
\textbf{HE Accuracy (SP$^{\mathrm{k}}$=64):} 75\% \newline 
\textbf{HE Accuracy (SP$^{\mathrm{k}}$=96):} 65\% \newline 
\textbf{HE Accuracy (SP$^{\mathrm{k}}$=128):} 60\% & 
\textbf{Secure Aggregation:} $\sim$80\% \newline 
\textbf{Partial Secure Aggregation:} $\sim$75\% \newline 
\textbf{Backdoor (5 rounds):} $\sim$0\% \newline 
\textbf{Backdoor (10 rounds):} $\sim$10\% \newline 
\textbf{Backdoor (60 rounds):} $\sim$50\% \newline 
\textbf{Backdoor (80 rounds):} $\sim$60\% \\
\hline
\end{tabular}
\begin{tabular}{l}
$^{\mathrm{a}}$FLPC: Federated Learning Parameter Compression, $^{\mathrm{b}}$mGAN-AI: Generative Adversarial Network for Adversarial Inference, \\
$^{\mathrm{c}}$PEFL: Privacy-Enhanced Federated Learning,
$^{\mathrm{d}}$DP: Differential Privacy, 
$^{\mathrm{e}}$CND: Clip Norm Decay, 
$^{\mathrm{f}}$PBA: Privacy Robust Aggregation,\\
$^{\mathrm{g}}$GA: Gaussian Attack, 
$^{\mathrm{h}}$LFA: Label Flipping Attack, 
$^{\mathrm{j}}$PB: Privacy Budget, 
$^{\mathrm{k}}$SP: Security Parameter, 
$^{\mathrm{i}}$HE: Homomorphic Encryption 
\end{tabular}
\vspace{-0.4cm}
\label{tab:quantitative_summary}
\end{table*}
\subsubsection{Encryption and Obfuscation}
These measures encrypt or alter data to prevent direct access or interpretation by unauthorised parties. \textbf{Gradient obfuscation} conceals sensitive data by altering gradient samples within FL processes to prevent direct inference attacks without sacrificing model performance. It protects data by making it difficult to reverse-engineer or identify sensitive information from gradients \cite{Wu2022Defense}. Yue et al. \cite{Yue2022} present an analysis of how gradient obfuscation, including quantisation and perturbation, provides a false sense of security in FL by demonstrating the feasibility of data reconstruction attacks despite these privacy measures. Fu et al. \cite{Fu2020} propose VFL, a verifiable FL framework for big data in the IIoT, enhancing privacy through Lagrange interpolation and blinding technology to safeguard gradient privacy. Gade et al. \cite{Gade2018} introduce a privacy-preserving distributed learning method using obfuscated stochastic gradients to enhance privacy against honest-but-curious adversaries in an FL setup. 

\textbf{Parameter compression} reduces detailed information sharing in FL, preventing attackers from reconstructing private data from model parameters. Zhu et al. \cite{Zhu2023Defending} address privacy inference attacks in FL for IoT via parameter compression, preserving privacy and model accuracy. Chen et al. \cite{Chen2023Efficient} discuss an adaptive federated optimisation algorithm that balances computation, communication, and precision in IoT environments using parameter compression.

\textbf{Compressed sensing as encryption} uses compressed sensing as a dual method for data compression and encryption, safeguarding gradients and labels against inference attacks. Miao et al. \cite{Miao2023Efficient} design an efficient privacy-preserving FL scheme based on compressed sensing, which serves both as a compression and encryption method. This approach ensures that gradients do not disclose private information, making it suitable for IoT scenarios. Li et al. \cite{Li2021Communication} propose FL algorithms based on compressed sensing, enhancing communication efficiency in IoT environments. These algorithms allow for model updates between IoT clients and a central server, improving performance over traditional methods.

\subsubsection{Differential Privacy and Noise Injection}
These measures use noise to mask data, adhering to differential privacy standards to ensure individual data points remain indiscernible. \textbf{Differential privacy-injected noise} incorporates artificial noise based on differential privacy to protect local parameters, balancing privacy with model accuracy. Shen et al. \cite{Shen2022Performance} have developed a performance-enhanced DP-based FL algorithm for IoT, introducing a classifier-perturbation regularisation method to improve the robustness of the trained model against DP-injected noise. Cui et al. \cite{Cui2021Security} have designed an improved differentially private FL system for anomaly detection in IoT infrastructures, optimising data utility throughout the training process. Ruzafa-Alcazar et al. \cite{Ruzafa-Alcazar2023Intrusion} provide a comprehensive evaluation of differential privacy techniques in the training of an FL-enabled intrusion detection system for IIoT. Yin et al. \cite{Yin2021Privacy} propose a new hybrid privacy-preserving method for federal learning that employs sparse differential gradient to improve transmission efficiency in social IoT scenarios. He et al. \cite{He2024Clustered} introduce adaptive local differential privacy mechanisms in FL for heterogeneous IoT data, focusing on balancing the trade-off between privacy and utility.While differential privacy techniques have shown promising results in controlled environments, their practical application in real-world IoT scenarios often faces challenges such as maintaining utility while ensuring privacy. Recent studies \cite{he2024adaptive} and \cite{Cui2021Security} have highlighted the need for adaptive mechanisms that balance this trade-off effectively.

\textbf{Decentralised perturbation techniques} distribute the task of injecting noise across federated nodes to protect privacy, enhancing the scalability and robustness of privacy measures~\cite{Arachchige2020Privacy}. Mothukuri et al.~\cite{Mothukuri2021Federated} propose an FL-based anomaly detection for IoT security, utilising decentralised data processing to enhance privacy and model accuracy. Mantey et al.~\cite{Mantey2023Blockchain} introduce a Secure Recommendation and Training Technique (SERTT) that leverages both FL and blockchain for privacy-preserved data management in the Internet of Medical Things (IoMT). Alotaibi \cite{Alotaibi2021Biserial} proposes a biserial correlative Miyaguchi–Preneel blockchain-based Ruzicka-indexed deep multi-layer perceptive learning (BCMPB-RIDMPL) method for improving malware detection in IoMT. Alamleh et al. \cite{Alamleh2022Federated} have developed a standardisation and bench-marking framework for machine-learning based intrusion detection systems using FL in IoMT environments.

\subsubsection{Secure Multi-party Computation and Anonymisation}
These measures focus on collaborative techniques that enable secure and private computations among multiple parties without revealing individual data inputs. \textbf{Secure multiparty computing} employs secure multiparty computing to enable private information exchange between FL participants, enhancing data privacy through complex protocols \cite{Li2020Privacy-Preserving}. Liu et al. \cite{Liu2023Comprehensive} propose a privacy-preserving FL scheme for Internet of Medical Things, which includes secure authentication and aggregation to protect data during model training. Lu et al. \cite{Lu2020Blockchain} have designed a blockchain and FL-based architecture for secure data sharing in IIoT, maintaining data privacy by sharing the data model instead of the actual data. Zhou et al. \cite{Zhou2020PrivacyPreserving} present an FL scheme in fog computing that enhances privacy and efficiency by integrating secure multi-party computing techniques. \textbf{Anonymisation and Siamese networks} use anonymisation strategies along with advanced network architectures to protect client identity and data during the training process, making re-identification challenging. Song et al. \cite{Song2020Analyzing} propose a framework incorporating GANs with a multi-task discriminator to analyse user-level privacy leakage in FL, developing a siamese network to re-identify anonymised updates and measuring the similarity of representatives effectively in IoT scenarios.

\section{Discussion}\label{Discussion}
This comprehensive review systematically explores the landscape of privacy threats in FL within IoT environments and evaluates the effectiveness of various defensive measures. We identify common threats such as inference and poisoning attacks and discuss lesser-covered threats such as Sybil and backdoor attacks in the context of IoT devices. The defensive measure of Differential Privacy is prominently featured, highlighting its critical role across various phases of the FL process. Additionally, we extend the current understanding by contrasting our findings with previous reviews which often focus on a narrower range of threats or do not address the unique challenges posed by the IoT environment. Notable papers by Mothukuri et al. \cite{Mothukuri2021A} and Ferrag et al. \cite{Ferrag2021Federated} primarily highlight security aspects without delving into the nuanced impacts of these environments on privacy and security strategies.

\subsection{Current Landscape of Threats and Defences}
In the reviewed papers, inference attacks are extensively studied, while gradient leakage and reconstruction are notably less addressed, indicating significant research gaps within federated learning in IoT (see Table~\ref{attack_citations}). Replay, evasion, and model stealing attacks also emerge as critical yet under-researched threats. The lack of focus on these vulnerabilities is concerning due to their potential to disrupt federated models' integrity and effectiveness. We emphasise the need for IoT system designers to incorporate robust defences early in the design phase. Defensive strategies such as differential privacy and secure multi-party computation, though promising, must be tailored to IoT constraints such as limited computational power and energy resources. Addressing these gaps is crucial for safeguarding systems against sophisticated cyber threats, ensuring reliability and trustworthiness in applications such as autonomous driving and medical diagnostics. In practical applications, secure multiparty computation in IoT has shown varying success. For instance, Liu et al. \cite{Liu2023Comprehensive} demonstrate its feasibility in medical IoT systems but noted significant computational overhead. Similarly, Lu et al. \cite{Lu2020Blockchain} highlight integrating blockchain with FL to enhance data integrity and privacy in IIoT, though it requires substantial computational resources that may not be available in all IoT settings.

\subsection{Advances and Innovations}
There is a critical need for developing lightweight privacy-preserving algorithms optimised for the IoT contexts. Our findings suggest that while differential privacy offers a balanced approach to privacy and efficiency, secure multi-party computation and other high-cost measures may require significant optimisation to be feasible in IoT contexts. Furthermore, emerging technologies like blockchain could offer scalable solutions but need thorough evaluation in real-world IoT settings to determine their operational viability. Additionally, empirical studies assessing the real-world applicability and resilience of proposed defensive mechanisms under varied IoT conditions and attack scenarios would greatly benefit the field. The expansion of IoT devices in sensitive areas (such as healthcare and smart cities) underscores the urgency of addressing privacy in FL. As IoT devices become more pervasive, ensuring the privacy and security of FL systems will be crucial in maintaining user trust and regulatory compliance.

\subsection{Limitations}
There are several limitations to our research, starting with the exclusion of gray literature and non-English publications, which might contain relevant data and insights. Additionally, the rapid evolution of both threats and technologies in this domain means that our findings might require continuous updates to remain relevant.

\section{Conclusion}\label{Conclusion}
In conclusion, this systematic review critically assesses privacy threats and defensive measures in Federated Learning (FL) within IoT environments. Analysing literature from 2017 to April 2024, we identified persistent challenges such as inference and poisoning attacks that compromise FL model robustness. The review highlights the need for innovative defensive strategies tailored to IoT constraints, balancing computational efficiency with privacy safeguards. Our findings emphasise integrating advanced measures such as Differential Privacy and Secure Multi-Party Computation to mitigate privacy risks. However, under-explored threats -- such as replay, evasion, and model stealing attacks -- pose significant risks, necessitating further research. 
Practical implementation of defensive measures in IoT settings reveals some potential but also exposes gaps requiring further research. Effective deployment demands addressing computational constraints and ensuring robust performance under variable network conditions, as shown in recent studies \cite{he2024adaptive, Liu2023Comprehensive}. Future research should prioritise developing lightweight, optimised privacy-preserving algorithms and explore emerging technologies such as  blockchain to enhance FL privacy. Additionally, developing FL models that operate under variable network conditions while maintaining edge device privacy is crucial. Further exploration into FL adaptations for edge computing to reduce latency and improve response times in privacy-critical applications is also essential.

\section*{Acknowledgements}
This work was supported by the funding received from the UK EPSRC project EP/X036707/1 on Countering HArms caused by Ransomware in the Internet Of Things (CHARIOT). The authors would also like to thank the anonymous reviewers for their constructive feedback.

\bibliography{refs}\label{References}
\bibliographystyle{ieeetr}
\end{document}